\documentclass[manuscript]{aastex}
\usepackage{graphicx}
\usepackage{txfonts}
\usepackage{natbib}

\shorttitle{Analytical Solutions for  Radiation-Driven Winds.}
\shortauthors{Araya et al.}

\begin{document}
\title{Analytical Solutions for Radiation-Driven Winds in Massive Stars. I: The Fast Regime.}
\author{I. Araya and M. Cur\'e} 
\affil{Instituto de F\'{\i}sica y Astronom\'{\i}a, Facultad de Ciencias, Universidad de Valpara\'{\i}so\\
Av. Gran Breta\~na 1111, Casilla 5030, Valpara\'{\i}so, Chile}
\email{ignacio.araya@uv.cl}
\and
\author{L. S. Cidale\altaffilmark{1}}
 \affil{Departamento de Espectroscop\'{\i}a, Facultad de Ciencias Astron\'omicas y Geof\'{\i}sicas, 
 Universidad Nacional de La Plata (UNLP), 
and\\ Instituto de Astrof\'{\i}sica La Plata, CCT La Plata, CONICET-UNLP\\
Paseo del Bosque S/N, 1900 La Plata, Argentina}

\altaffiltext{1}{Member of the Carrera del Investigador Cient\'{\i}fico, CONICET, Argentina}


 \begin{abstract}
 Accurate mass--loss rate estimates are crucial keys to study wind properties of massive stars and test different evolutionary scenarios. From a theoretical point of view, this implies to solve a complex set of differential equations in which the radiation field and the hydrodynamics are strong coupled. The use of analytical expression to represent the radiation force and the solution of the equation of motion have many advantages over numerical integrations. 
Therefore, in this work, we present an analytical expression as solution of the equation of motion for radiation--driven winds, in terms of the force multipliers parameters. This analytical expression is obtained by employing the line acceleration expression given by \citet{Villata1992} and the methodology proposed by \citet{Muller2008}. On the other hand, we find useful relationships to determine the parameters for the line acceleration given by  \citet{Muller2008} in terms of the force multiplier parameters.
\end{abstract}

%
%
\section{Introduction}

The study of the winds of massive stars is very important in many aspects. They affect not only the nearby environment of the stars, through the input of energy and momentum, but also the evolution of the mass--losing star itself. The interaction of the expelled outflow with the interstellar medium (ISM) contributes to the formation of majestic nebulae,  stellar filaments, bow--shock, blown--bubbles, bipolar jets, and circumstellar disks/rings. All these fascinating structures are far from being spherical and homogeneous. 

Winds of massive stars are mainly driven by the line--radiation force \citep[CAK and m-CAK theories, see][respectively]{castor1975,ppk1986}. Moreover, nowadays, there is more evidence that radiation--driven winds are unstable and highly variable \citep[see e.g.,][]{puls2008}.  Numerical simulations of time-dependent models show that the non--linear evolution of wind  instabilities leads to the formation of shocks and spatial structures in both density and velocity \citep[named clumps,][]{Dessart2005}. The development of clumps due to 
the line--deshadowing instability (LDI) is expected when the wind velocity is large enough to produce shocked structures.  However, new observations indicate that clumping is already present close to the
stellar photosphere. Such inhomogeneities could be then related to waves produced by the sub-surface convection zone \citep{cantiello2009}. In addition, any notable inhomogeneity will necessarily result in a mass--loss rate overestimate. Therefore, a clumped--wind theory is essential to resolve the mass--loss discrepancy \citep{surlan2013}, but to avoid time expensive computations analytical expressions are required.

During the star evolution, evolved massive stars transient short-lived phases (i.e.; blue
supergiants, Luminous Blue Variables (LBVs), B[e] supergiants and RGB stars) and expel huge amounts of mass to the ISM.  The rate at which massive stars lose mass is quite uncertain and
still subject of debate. Therefore,  accurate mass--loss rate estimates are crucial keys to study the wind properties of O--B stars and test different evolutionary scenarios. The amount of mass lost via stellar winds can be evaluated  by comparing theoretical results with observation and studying their effects on the emergent line spectrum. From a theoretical point of view, this implies to solve a complex set of differential equations in which the radiation field and the hydrodynamics are strong coupled by the radiation force  and its interaction with the medium. To treat efficiently this system of equations, in most of the cases, the hydrodynamics solution is approximated  by a simple analytical expression, such as a $\beta$ velocity law \citep[see e.g.,][]{castor1979} with typical $\beta$ values determined empirically by modelling the observed line profiles. Furthermore, analytical solution of the hydrodynamics equations are indispensable  to deal with multidimensional radiative transfer problems for moving media, using Monte Carlo techniques with reasonable accuracy on time--scales of a few hours. Therefore, the velocity profile is frequently described with a $\beta$--law or a double $\beta$--law \citep{surlan2013,hillier1999}.

The scope of this work is to deliver an accurate and convenient analytical expression for the solution of the equation of motion for radiation--driven winds, by employing a line acceleration expression in terms of the force multiplier parameters ($\alpha$, $k$ and $\delta$). The success of these parameters is that they provide with scaling laws for the mass-loss rate and terminal speed as well as with an empirically observed “Wind-Momentum-Luminosity” (WML) relationship that depends on metallicity \citep{Owocki2009}. On the other hand, the CAK line  force multipliers $\alpha$ and $k$ describe the statistical dependence of the number of lines on frequency position and line-strength (e.g. Puls et al. 2000) while the parameter $\delta$ is related to the change in the ionization of the wind. CAK parameters have been improved using NLTE calculations and they are adequate to describe the optically thin winds of O and B stars.
The use of analytical expressions for the radiation force and velocity profile as function of the force multiplier parameters provides a clear view into how the line-driven mechanism affect the hydrodynamics and clumping. It also might help to treat problems with abrupt discontinuity in the stellar parameters of B supergiants, such as the bi--stability jump as consequence of changes in the wind ionization \citep{lamers1991,vink1999,cure2005}. 

An alternative approach to CAK involves the Monte Carlo method developed by \citet{Abbott1985} that leads  to a parametrised description of the line acceleration that only depends on radius (rather than explicitly on the velocity gradient dv/dr as in
the standard CAK theory.) The new parameters that fit the Monte Carlo line acceleration describe very well the dense winds of O stars \citep{vink2014} but they are not so directly related to the main mechanism driving the wind, such as line strength and  ionization. Therefore, we also provide useful relationships between the CAK line force parameters and the M\"uller\& Vink's (\citeyear{Muller2008}) parametrization, making them useful for thin and dense winds.

In a forthcoming paper, we plan to develop other analytical expressions adequate to describe radiation--driven  winds for stars rotating near the critical rotation velocity \citep[][named $\Omega$--slow solutions]{cure2004}, or radiation--driven winds with ionization gradients \citep[][$\delta$--slow solutions]{cure2011}.

In section \S \ref{basic-eq} we outline the main steps to obtain the dimensionless form of the equation of motion. In addition, we review the analytical expressions of the line accelerations given by \citet{Villata1992} and \citet{Muller2008}.
In section \S \ref{results}, we give our own analytical approximation of the line acceleration and discuss useful expressions that enable to obtain the parametrization of M\"uller \& Vink's (\citeyear{Muller2008}) line--acceleration in terms of the force multiplier parameters. 

Discussion on the results and conclusions are presented in sections \S \ref{discussion} and \S \ref{conclusions}, respectively.

\section{The standard hydrodynamical wind model}
\label{basic-eq}
The {\it CAK theory} for a radiation-driven wind was developed by \citet{castor1975} who describe a stationary, one dimensional, non--rotating, isothermal, outflowing wind in spherical symmetry. Using these hypotheses, and neglecting the effects of viscosity, heat conduction and  magnetic fields, the equations of mass conservation and radial momentum state:

\begin{equation}
\label{continuity}
4\, \pi \, r^{2}\, \rho \, v = \dot{M}, 
\end{equation}

\noindent and

\begin{equation}
\label{momentum}
v \, \frac{dv}{dr}=-\frac{1}{\rho}\frac{dp}{dr} - \frac{G\, M_{*} (1-\Gamma)}{r^{2}} + g^{\rm{line}}(\rho, dv/dr, n_{E}).
\end{equation}

\noindent Here $v$ is the fluid radial velocity, $dv/dr$ is the velocity gradient and g$^{line}$ is the radiative acceleration. All other variables have their standard meaning \citep[for a detailed derivation and definitions of variables, constants and functions, see][]{cure2004}. 

The CAK theory assumes that the radiation emerges directly from the star (as a source point) and multiple scatterings are not taken into account. A later improvement of this theory (m-CAK) considers the radiation coming from a stellar disk, and therefore, we adopt the m-CAK standard parametrization for the line force term, following the descriptions of  \citet{abbott1982}, \citet{friend1986} 
and \citet{ppk1986}, namely:

\begin{equation}
g^{\rm{line}} = \frac{C}{r^{2}}\, f_{D}(r,v,dv/dr)\, \left( r^{2} \, v \, \frac{dv}{dr} \right)^{\alpha} \left( \frac{n_{E}}{W(r)} \right)^{\delta},
\end{equation}

\noindent where the coefficient (eigenvalue) $C$ depends on the mass-loss rate, $\dot{M}$, $W(r)$ is the dilution factor, $n_{E}$ is the electron number density in units of $10^{-11}\, \rm{cm^{-3}}$ and $f_{D}$ is the finite disk correction factor.\\

The momentum equation (Eq. \ref{momentum}) can be expressed in a dimensionless form  \citep[see e.g.,][]{Muller2008} as, 

\begin{equation}
\hat{v} \, \frac{d \hat{v}}{d \hat{r}}= -\frac{\hat{v}_{\rm{crit}}^{2}}{\hat{r}^{2}} + \hat{g}^{\rm{line}} - \frac{1}{\rho}\frac{d \rho}{d \hat{r}},
\end{equation}

\noindent where $\hat{r}$ is a dimensionless radial coordinate $\hat{r}=r/R_{*}$, and the dimensionless velocities 
(in units of the isothermal sound speed $a$) are:

\begin{equation}
\hat{v}=\frac{v}{a}\:\:\:\:\: \mathrm{and} \:\:\:\:\: \hat{v}_{\rm{crit}}=\frac{v_{\rm{esc}}}{a\sqrt{2}},
\end{equation}

\noindent $v_{\rm{crit}}$ is the rotational break-up velocity in the case of a corresponding rotating star, but 
often it is simply defined by the effective escape velocity $v_{\rm{esc}}$ divided by a factor of $\sqrt{2}$. In the same way, we can write a dimensionless line acceleration by:

\begin{equation}
\label{norma}
\hat{g}^{\rm{line}}=\frac{R_{*}}{a^{2}}\, g^{\rm{line}}.
\end{equation}

\noindent With the equation of state of an ideal gas ($p=a^{2}\rho$) and using equation (\ref{continuity}), the dimensionless equation of motion is the following:

\begin{equation}
\label{motion}
\left( \hat{v} - \frac{1}{\hat{v}} \right) \frac{d\hat{v}}{d \hat{r}}= -\frac{\hat{v}_{\rm{crit}}^{2}}{\hat{r}^{2}} + \frac{2}{\hat{r}} + \hat{g}^{\rm{line}}.
\end{equation}
\vskip 0.3cm
\subsection{Line Acceleration}
\label{gline}
%
%
In this section we recapitulate the basic concepts developed by \citet{Villata1992} and \citet{Muller2008} to derive later a general analytical expressions for the velocity profile in the frame of the standard radiation-driven wind 
solution for massive stars. These basic concepts will then be used in a forthcoming work to obtain analytical expressions for the $\Omega$-- and $\delta$--slow solutions.  

For that purpose we analyse two expressions for the line acceleration which are functions
only on the radial distance and do not depend neither on the velocity nor the velocity gradient as the standard m--CAK description does.
We also demonstrate that both expressions are related to each other and allow a noticeable simplification to integrate 
the equation of motion (Eq. \ref{motion}) leading to an analytical expression for the fast wind's velocity profile.

\subsubsection{Villata's approximation}
Based on the analytic study of radiation-driven stellar winds by \citet{kppa1989},  \citet{Villata1992} derived 
an approximated expression for the line acceleration term. 
This line acceleration depends only on the radial coordinate, and reads: \\

\begin{equation}
g^{line}_{\mathrm{V92}}(\hat{r})=\frac{G\,M_{*}\,(1-\Gamma)}{R_{*}^{2}\,\hat{r}^{2}}\,A(\alpha, \beta, \delta) \left( 1- \frac{1}{\hat{r}}\right)^{\alpha (2.2 \, \beta -1)},
\label{villata-gline}
\end{equation}

\noindent with
\begin{eqnarray}
\label{villata-gline2}
A(\alpha, \beta, \delta)=\frac{(1.76\, \beta)^{\alpha}}{1-\alpha}\left[10^{-\delta}(1+\alpha)\right]^{1/(1-\alpha)}  \nonumber \\
\left[ 1 + \left( \frac{2}{\alpha} \left\lbrace 1- \left[ 10^{-\delta} (1+\alpha)\right] ^{1/(\alpha -1)} \right\rbrace \right) ^{1/2} \right] ^{\alpha} ,
\end{eqnarray}

\noindent where $\alpha$ and $\delta$ are force multiplier parameters \citep{abbott1982}, and $\beta$ is the exponent in the $\beta$ velocity law. This exponent can be evaluated with a formula given by \citet{kpp1987} in terms of the force multiplier parameters and the escape velocity, $v_{\rm{esc}}$:

\begin{equation}
\beta = 0.95 \, \alpha + \frac{0.008}{\delta}+\frac{0.032 \, v_{\rm{esc}}}{500} , 
\end{equation}

\noindent with $v_{\rm{esc}}$ in km\,s$^{-1}$.

Then, using Eq. (\ref{villata-gline}) in its dimensionless form (Eq. \ref{norma}) and inserting it into the  dimensionless equation of motion (Eq. \ref{motion}), it yields:

\begin{equation}
\label{villata-motion}
\left( \hat{v} - \frac{1}{\hat{v}} \right) \frac{d\hat{v}}{d \hat{r}} = -\frac{\hat{v}_{\rm{crit}}^{2}}{\hat{r}^{2}} + \frac{2}{\hat{r}}  
+ \frac{1}{a^{2}} \frac{GM_{*} (1-\Gamma)}{R_{*}\,\hat{r}^{2}}\, A(\alpha, \beta, \delta) \left( 1- \frac{1}{\hat{r}}\right)^{\gamma_{\rm v}},
\end{equation}
\\

\noindent with $\gamma_{\rm v}\,=\,{\alpha \, (2.2 \, \beta -1)}.$

This differential equation, based on the approximation of the line acceleration given by Villata, $\hat{g}^{line}_{\mathrm{V92}}(\hat{r})$, is a solar-like differential equation of motion. In particular, the expression  $\hat{g}^{line}_{\mathrm{V92}}(\hat{r})$ does not 
depend on the product $\hat{v} \, d\hat{v}/d\hat{r}$, as in the case of the standard m--CAK theory, and hence the singular point is the sonic point, e.g., when the velocity equals the sound speed.
Another important difference between Villata's equation of motion and its equivalent equation in the standard m--CAK theory is 
that Eq. (\ref{villata-motion}) has no eigenvalues. This means that the differential equation does not depend explicitly on the 
mass--loss rate ($\dot{M}$) of the star. Therefore, Villata derived the following expression for $\dot{M}$:

\begin{equation}
\label{mdot-villata}
\dot{M}= 1.2 \, \left( \frac{D^{\delta}\, \dot{M}^{\alpha}_{\mathrm{CAK}}}{1+\alpha} \right)^{1/(\alpha - \delta)},
\end{equation}

\noindent where $D$ and $\dot{M}_{\mathrm{CAK}}$ are given by

\begin{equation}
D= \left( \frac{1 + Z_{\rm{He}} \, Y_{\rm{He}}}{1 + 4 \, Y_{\rm{He}}} \right) \left( \frac{9.5 \times 10^{-11}}{\pi \, m_{\rm{H}} \, R^{2}_{*} \, v_{\infty}} \right)
\end{equation}

\noindent and

\begin{equation}
\label{mcak}
\dot{M}_{\mathrm{CAK}}= \frac{4\, \pi \, G \, M_{*} \, \alpha}{\sigma_{\rm{E}}\, v_{\rm{th}}} \left[ k \, \Gamma \left( \frac{1-\alpha}{1-\Gamma} \right)^{1-\alpha} \right]^{1/\alpha},
\end{equation}

\noindent where $Z_{\rm{He}}$ is the amount of free electrons provided by helium, $Y_{\rm{He}}$ is the helium abundance relative to hydrogen, $m_{\rm H}$ is the proton mass, $\sigma_{\rm E}$ is the Thomson scattering absorption coefficient per mass density, $v_{\infty}$ is the wind terminal velocity, $v_{\rm{th}}$ is the thermal velocity of protons and $k$ is a force multiplier parameter; all these parameters are given in cgs units. 

Villata solved the equation of motion only by means of a standard numerical integration, obtaining terminal velocities that agree, within 3--4 $\%$, with those
computed by \citet{ppk1986} and \citet{kpp1987}. In addition, the mass--loss rate he obtained agrees very well with the numerical results computed by \citet{ppk1986} \citep[see details in][Table 1]{Villata1992}.

\subsubsection{M\"uller \& Vink's approximation}

In the context of stellar wind theory of massive stars, \citet{Muller2008} (hereafter MV08) present an analytical 
expression for the velocity field using a parametrised description for the line acceleration that (as in Villata's) it depends also on the radial coordinate.  \citet{Muller2008} computed the line acceleration using Monte-Carlo multi-line radiative transfer calculations  
\citep{koter1997,vink1999} and a $\beta$ velocity law. Then, the numerical line acceleration was fitted by the following function:

\begin{equation}
\label{MV-gline}
\hat{g}^{line}_{\mathrm{MV08}}(\hat{r})= \frac{\hat{g}_{0}}{\hat{r}^{1+ \delta_{1}}} \left(  1-\frac{\hat{r_{0}}}{\hat{r}^{\delta_{1}}} \right) ^{\gamma},
\end{equation}

\noindent where $\hat{g_{0}}$, $\delta_{1}$, $\hat{r_{0}}$ and $\gamma$ are acceleration line parameters.

Replacing Eq. (\ref{MV-gline}) in Eq. (\ref{motion}), MV08 derived the following dimensionless equation of motion,

\begin{equation}
\label{MV-motion}
\left( \hat{v} - \frac{1}{\hat{v}} \right) \frac{d\hat{v}}{d \hat{r}}=
-\frac{\hat{v}_{crit}^{2}}{\hat{r}^{2}} + \frac{2}{\hat{r}} 
+ \frac{\hat{g}_{0}}{\hat{r}^{1+ \delta_{1}}} \left(  1-\frac{\hat{r_{0}}}{\hat{r}^{\delta_{1}}} \right) ^{\gamma},
\end{equation}

\noindent and a fully {\it{analytical}} 1-D velocity profile \citep[see][for details about the methodology used to obtain this solution]{Muller2008} as function of the so-called  Lambert W--function \citep{corless1993,corless1996,cranmer2004} . The analytical solution reads: 

\begin{equation}
\label{V-profile}
\hat{v}(\hat{r})= \sqrt{-W_{j}(x(\hat{r}))},
\end{equation}

\noindent with 

\begin{equation}
\label{eq-X}
x(\hat{r})= -\left(  \frac{\hat{r}_{\rm c}}{\hat{r}} \right) ^{4}  \; \exp \left[ -  2 \, \hat{v}^{2}_{\rm{crit}} \left( \frac{1}{\hat{r}} - \frac{1}{\hat{r}_{\rm c}}  \right)
\\
-\frac{2}{\hat{r}_{0}} \frac{\hat{g}_{0}}{\delta_{1} \, (1+ \gamma)}  \left[ \left( 1- \frac{\hat{r}_{0}}{\hat{r}^{\delta_{1}}} \right) ^{1+\gamma} - \left( 1- \frac{\hat{r}_{0}}{\hat{r}_{\rm c}^{\delta_{1}}} \right) ^{1+\gamma} \right]  - 1 \right].
\end{equation}

\noindent In the last equation appears the new parameter $\hat{r}_{\rm c}$, which represents the position of the sonic (or critical) point. 
Furthermore, it is important to notice that the Lambert W--function ($W_j$) has only 2 real branches,  which are denoted by a 
sub-index {\it j}, with $j=-1$ or $0$. The sonic point ($\hat{r_{\rm c}}$) is the boundary between 
these 2 branches, i.e.,

\begin{equation}
j = \left\{  \begin{array}{r r c}
    0 & \mathrm{for} &  1\leq \hat{r} \leq \hat{r}_{\rm c}\\
    -1 & \mathrm{for} & \hat{r}>\hat{r}_{\rm c},
  \end{array} \right. 
\end{equation}

On the other hand, we can observe that the LHS of
the equation of motion (Eq. \ref{MV-motion}) vanishes when $\hat{v}=1$ (singularity condition in the CAK formalism), therefore, as in the m-CAK case, a regularity condition must be imposed, which is equivalent to set that the RHS of Eq. (\ref{MV-motion}) also vanishes at $\hat{r}=\hat{r}_{\rm c}$. That is, 
\begin{equation}
 -\frac{\hat{v}_{\rm{crit}}^{2}}{\hat{r}_{\rm c}^{2}} + \frac{2}{\hat{r}_{\rm c}} + \frac{\hat{g}_{0}}{\hat{r}_{\rm c}^{1+ \delta_{1}}} \left(  1-\frac{\hat{r}_{0}}{\hat{r}_{\rm c}^{\delta_{1}}} \right) ^{\gamma}=0 ,
\end{equation}
\noindent and $\hat{r}_{\rm c}$ is obtained by solving numerically this last equation for a given set of the parameters, $\hat{g_{0}}$, $\delta_{1}$, $\hat{r_{0}}$ and $\gamma$. Then, the function $x(\hat{r})$ (Eq. \ref{eq-X}) is calculated and the velocity profile is derived  by replacing it in Eq. (\ref{V-profile}).

\section{Results}
\label{results}

The approximations described before have the advantages of providing analytical expressions to represent the radiation force. This fact simplifies considerably the resolution of the equation of motion.
However, each one of the mentioned approximations has its own advantages and disadvantages. Even when Villata's approximation of the radiation force is general and can be directly applied to describe the wind of any massive star, one still needs to deal with a numerical integration to solve the momentum equation\footnote{The definition of Lambert W--function was 1 year after Villata's work.}. The M\"uller \& Vink's approximation got an analytical solution of the equation of motion (through the Lambert W--function), in terms of $\hat{g_{0}}$, $\delta_{1}$, $\hat{r_{0}}$ and $\gamma$ parameters of the star. However, to obtain these parameters the line acceleration still needs to be derived using Monte-Carlo multi-line radiative transfer calculations.

Therefore, here, we propose to achieve a complete analytical description of the 1-D hydrodynamical solution for radiation-driven winds  gathering the advantages of both previous approximations (the use of known parameters and the Lambert W--function).

\subsection{Analytical solution in terms of the force multiplier and stellar parameters}

Our goal is to derive a new analytical expression combining Villata's expression of the equation of motion  (Eq. \ref{villata-motion}), based on Villata's definition of the radiation force, $\hat{g}^{\rm{line}}_{\mathrm{V92}}(\hat{r})$, with 
the methodology developed by MV08 to solve the equation of motion.
This way, we solve Eq. (\ref{villata-motion}) through the Lambert W--function \citep{corless1993,corless1996,cranmer2004}, 

 
\begin{equation}
\label{sol-V-MV}
\hat{v}(\hat{r})= \sqrt{-W_{j}(x(\hat{r}))},
\end{equation}

\noindent with

\begin{equation}
x(\hat{r})= -\left(\frac{\hat{r}_{\rm c}}{\hat{r}} \right)^{4} \, \exp\left[-  2 \,  \hat{v}^{2}_{\rm{crit}} \left( \frac{1}{\hat{r}} - \frac{1}{\hat{r}_{\rm c}}  \right)
- 2 \left(  I_{\hat{g}_{V92}^{\rm{line}}}(\hat{r}) -  I_{\hat{g}_{V92}^{\rm{line}}}(\hat{r}_{\rm c})   \right)  - 1 \right],
\end{equation}

\noindent where 

\begin{eqnarray}
I_{\hat{g}_{V92}^{\rm{line}}}(\hat{r})  = \left(10^{-\delta} \, (1+\alpha)\right)^{\frac{1}{1-\alpha}} \, \left(1+\sqrt{2} \sqrt{-\frac{\left(10^{-\delta} \, (1+\alpha)-1\right)^{\frac{1}{\alpha -1}}}{\alpha}}\right)^{\alpha} \nonumber \\
 (1.76 \, \beta)^{\alpha} \, G\,M_{*} \left(\frac{\hat{r}-1}{\hat{r}}\right)^{1+\gamma_{\rm v}}   \frac{\Gamma -1}{\left( a^2 [\alpha -1] (1+\gamma_{\rm v}) \, R_{*} \right) }.
\end{eqnarray}

\noindent  From the singular and regular condition at the critical or some point, $\hat{r}_{\rm c}$ can be obtained numerically making the RHS of Eq. (\ref{villata-motion}) equal zero, that is:

\begin{equation}
 -\frac{\hat{v}_{\rm {crit}}^{2}}{\hat{r}_{\rm c}^{2}} + \frac{2}{\hat{r}_{\rm c}} +  \frac{1}{a^{2}} \frac{G\,M_{*}\, (1-\Gamma)}{R_{*} \, \hat{r}_{\rm c}^{2}} \, A(\alpha, \beta, \delta) \left( 1- \frac{1}{\hat{r}_{\rm c}}\right)^{\gamma_{\rm v}} = 0.
\end{equation}


The advantages of  Eq. (\ref{sol-V-MV}) is that not only is based on the Lambert W-function but also depends on the force multiplier parameters and the stellar parameters.  Both stellar and force multiplier parameters are given for a wide range of spectral types \citep[see for example][]{abbott1982,ppk1986,lamers1999}.

%
%

\subsection{Comparison of the new general analytical expression with the solution for radiation--driven winds}
\label{compare}

The accuracy of the {\it new} analytical solution has to be tested by comparing it with rigorous numerical 1--D hydrodynamical solutions for radiation-driven winds, as described by \citet{cure2004}. 
Table \ref{sample} gives the values of $v_{\infty}$ obtained from Eq. (\ref{sol-V-MV}), together with those values calculated from  numerical results using the same sample of stars listed by \citet{Villata1992}. $v_{\infty}$ was calculated at 100 R$_{*}$ from the star, and $\dot{M}$ with the Eq. (\ref{mdot-villata}) derived by Villata but employing our $v_{\infty}$ estimates.

\begin{figure}[h]
\center
\includegraphics[width=4 in]{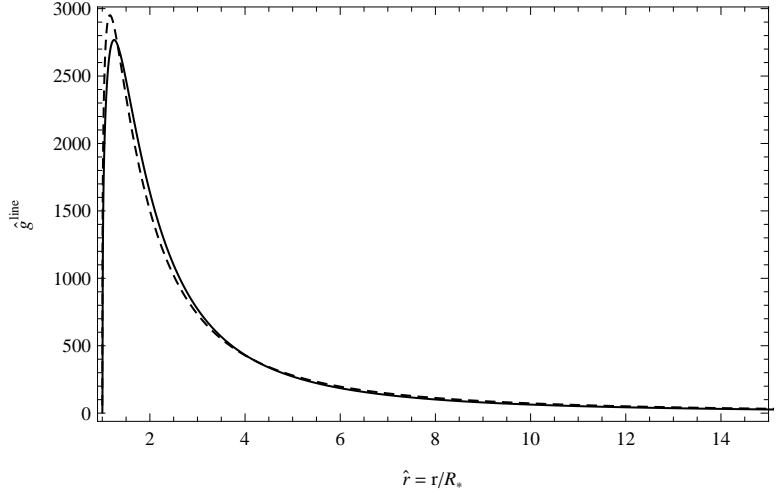}
\caption{Line acceleration as function of the radial coordinate $\hat{r}=r/R_{*}$. The hydrodynamic line acceleration is shown in solid line and $\hat{g}^{\rm{line}}_{V92}$ is in dashed--line. The calculation is based on a  set of values for $\epsilon$ Ori (see Table \ref{sample}). 
\label{gline-plot}}
\end{figure}	

\begin{figure}[h]
\center
\includegraphics[width=4 in]{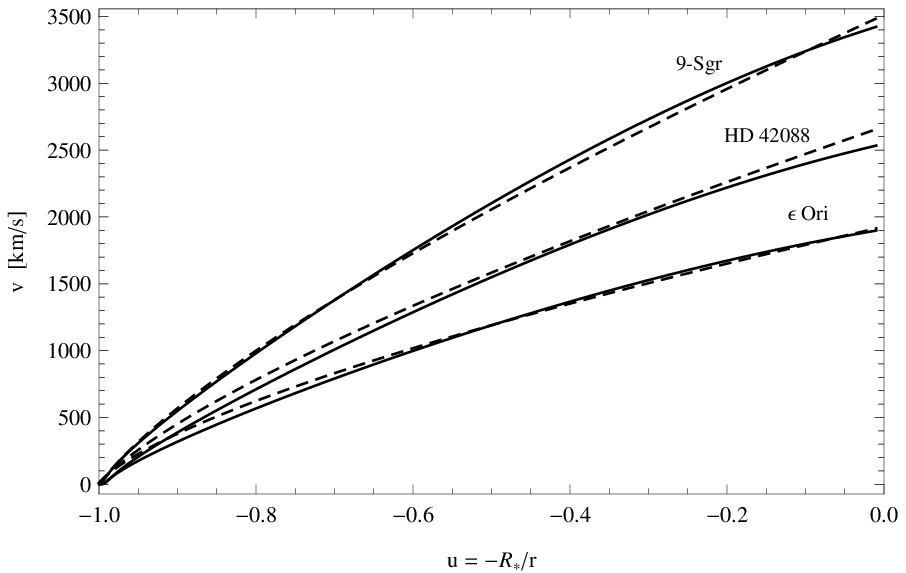}\\
\vspace*{0.4in}
\includegraphics[width=4 in]{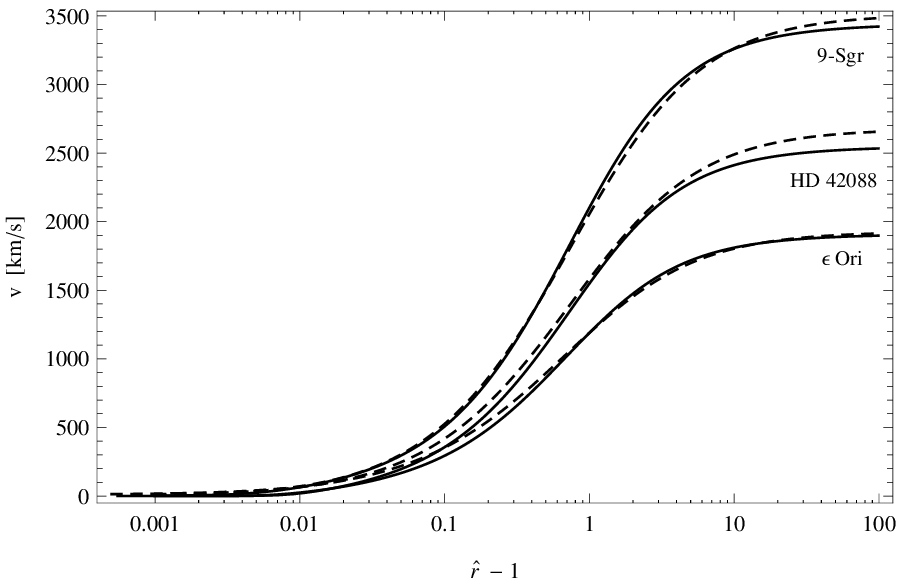}
\caption{Velocity profiles as function of the inverse radial coordinate $u=-R_{*}/r=-1/\hat{r}$ (upper panel) and $\hat{r}\,-\,1$ in logarithmic scale (lower panel) for three stars of the sample. The hydrodynamic results are shown in solid line and the analytical solutions are in dashed--line.
\label{profiles}}
\end{figure}

\begin{figure}[h]
\center
\includegraphics[width=4 in]{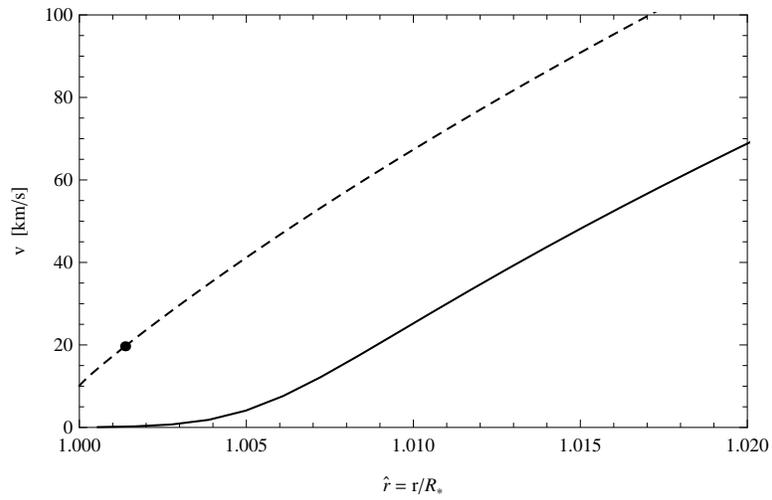}
\caption{Velocity profile as function of the radial coordinate $\hat{r}=r/R_{*}$ in a region very close to the surface of the star ($\epsilon$ Ori). The numerical hydrodynamic result is shown in solid line and the analytical solution is in dashed--line. The dot symbol indicates the position of the sonic (or critical) radius. 
\label{profile-zoom}}
\end{figure}

\begin{figure}[h]
\center
\includegraphics[width=4 in]{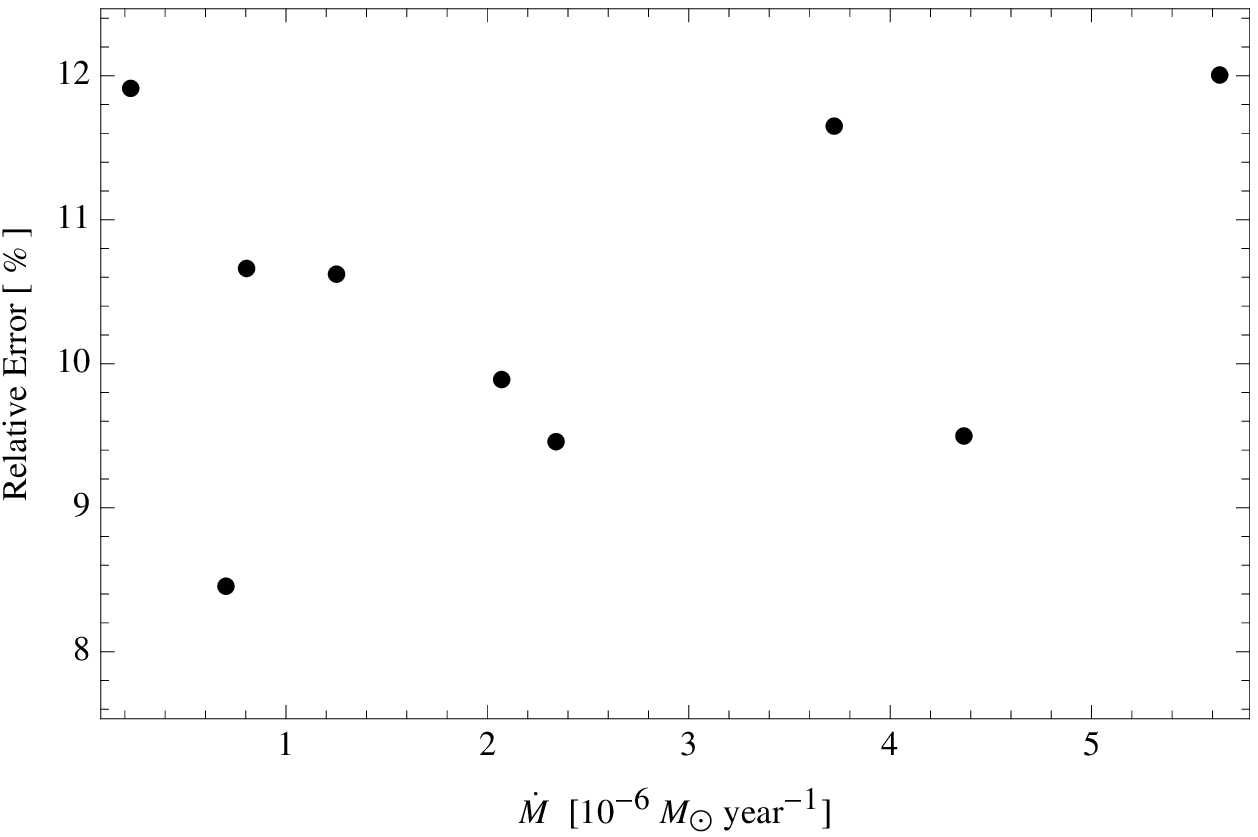}\\
\vspace*{0.4in}
\includegraphics[width=4 in]{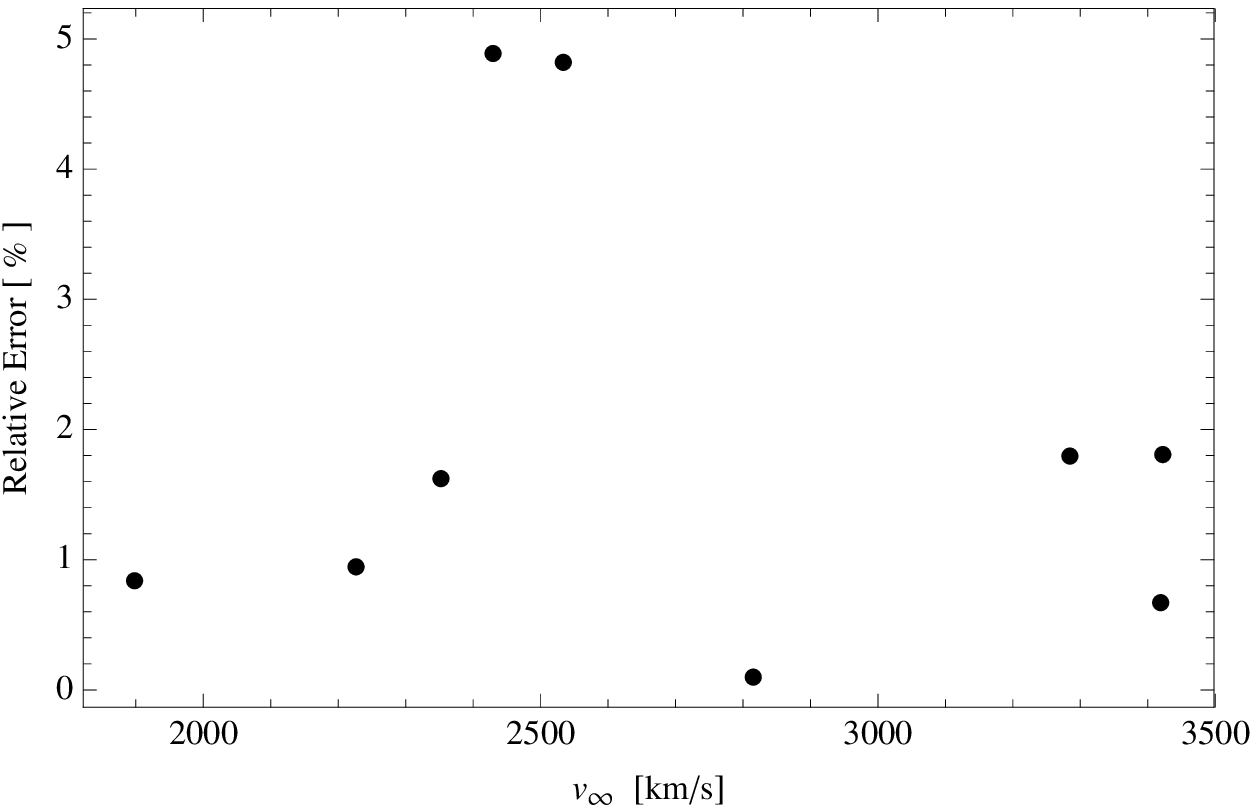}
\caption{Relative error for the mass--loss rate (upper panel) and terminal velocity (lower panel) respect to the values obtained from the hydrodynamics. The relative error for the terminal velocity has a median of $1.6\,\%$  and a mean of $1.9\,\%$. The median of the mass--loss rate is $10.6\,\%$ and the mean is $10.5\,\%$.}
\label{error}
\end{figure}

As an example, Fig. \ref{gline-plot} shows the good agreement found between the line acceleration obtained for $\epsilon$ Ori, using a numerical hydrodynamic model and the expression $\hat{g}^{\rm{line}}_{V92}$. 
The comparison of the resulting velocity profiles obtained via the hydrodynamics and the analytical solution of this work is shown in Fig. \ref{profiles} for three stars of the sample ($\epsilon$ Ori, 9\,Sgr and HD\,42\,088). Both analytical and numerical solutions seem to behave in a similar way. However, if we have a close view to the region near the stellar surface, that is the base of the wind (Fig. \ref{profile-zoom}), we can note that the velocity law derived from  the analytical solution is steeper than the numerical one. 
 The accuracy of our solution is reflected in Fig. \ref{error} which shows the relative error for the mass--loss rate and terminal velocity obtained from our analytical solution and  the hydrodynamical code.

In addition, from Table \ref{sample}, we determine the average value of $\hat{r}_{c}$ ($\bar{\hat{r}_{c}}=1.0026$) and re-calculate $v_{\infty}$ and $\dot{M}$ for the sample of selected stars. The results obtained are almost the same to those shown in Table \ref{sample}. This simplifies by far the numerical calculation of $\hat{r}_{c}$, obtaining thus a totally analytical solution. It is important to remark that the use of $\bar{\hat{r}_{c}}$ only can be applied in the supersonic region of the velocity profile, primarily to obtain the terminal velocity,  because of its close connection with both branches of Lambert W--function, which are discontinuous when we use $\hat{r}_{c}$.

\begin{table*}[ht]
  \begin{center}
  \tabcolsep 2.0 pt
  \caption{Comparison of the wind parameters obtained via the new analytical solutions with numerical calculations. The sample of stars was taken from \citet{Villata1992}.}
  \vskip 0.5cm
  \label{sample}
 {\scriptsize
  \begin{tabular}{lccccccccccccc}
\hline
\hline
Star & $T_{\mathrm{eff}}$ & $\log\,g$ & $R_{*}$ & $k$ & $\alpha$ & $\delta$ & $\hat{r}_{c}$ & $v_{\infty}$ & $v_{\infty}^{\mathrm{{\tiny Analytic}}}$ & $\dot{M}$ & $\dot{M}^{\mathrm{{\tiny Analytic}}}$\\
 &[$10^{3}$ K]&&[R$_{\odot}$]&&&&&[km\,s$^{-1}$]&[km\,s$^{-1}$]&[$10^{-6}\, M_{\odot}$\,yr$^{-1}$]&[$10^{-6}\, M_{\odot}$\,yr$^{-1}$]\\
\hline

$\epsilon$ Ori & 28.5 &  3.25 & 37 & 0.170 & 0.590 & 0.090&1.0014 &  1905 & 1914 & 3.71 & 3.29\\
$\zeta$ OriA & 30.0 &  3.45 & 29 & 0.170 & 0.590 & 0.090& 1.0017 & 2226  & 2205 & 2.07 & 1.87\\
9-Sgr & 50.0 &  4.08 & 12 & 0.124 & 0.640 & 0.070&1.0035  & 3422  & 3484 & 4.37 & 3.95\\
HD 48099 & 39.0 &  4.00 & 11 & 0.124 & 0.640 & 0.070& 1.0035 & 3419  & 3442 & 0.70  & 0.64\\
HD 42088 & 40.0 &  4.05 & 5.8 & 0.124 & 0.640 & 0.070&1.0027  & 2534  & 2656 & 0.23 & 0.20\\
$\lambda$ Cep & 42.0 &  3.70 & 17 & 0.124 & 0.640 & 0.070&1.0026  & 2430  & 2548 & 5.64 & 4.96\\
(O5 V)$_{\mathrm{Gal}}$ & 45.0 &  4.00 & 12 & 0.124 & 0.640 & 0.070&1.0034  & 3284  & 3343 & 2.34 & 2.12\\
(O5 V)$_{\mathrm{LMC}}$ & 45.0 &  4.00 & 12 & 0.089 & 0.627 & 0.100&1.0029  & 2815  & 2812 & 1.25 & 1.12\\
(O5 V)$_{\mathrm{SMC}}$ & 45.0 &  4.00 & 12 & 0.097 & 0.580 & 0.104& 1.0019 & 2352  & 2314 & 0.80 & 0.72\\

\hline
  \end{tabular}
  }
 \end{center}
\end{table*}

\section{Discussion}
\label{discussion}

\subsection{On the previous known analytical expressions.}

\begin{figure}[ht]
\center
\includegraphics[width=4 in]{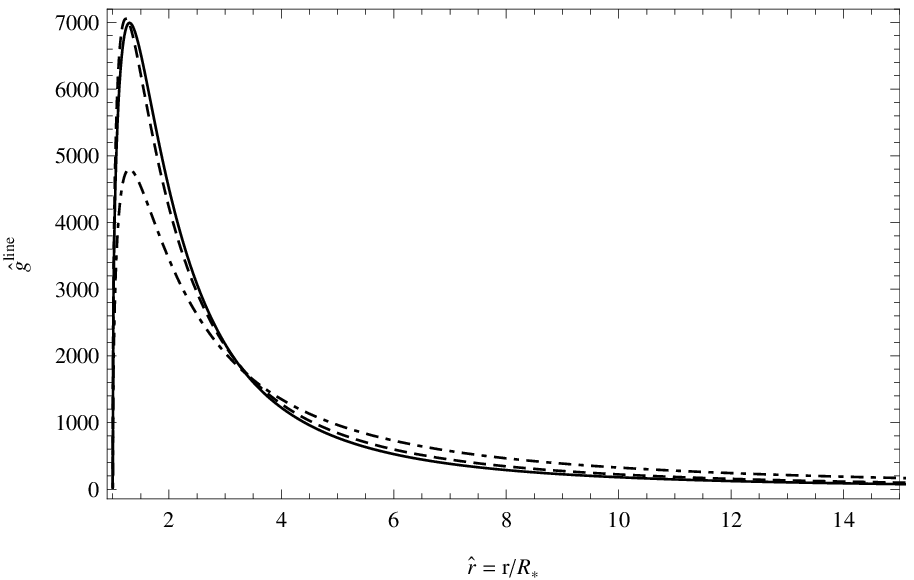}\\
\vspace*{0.4in}
\includegraphics[width=4 in]{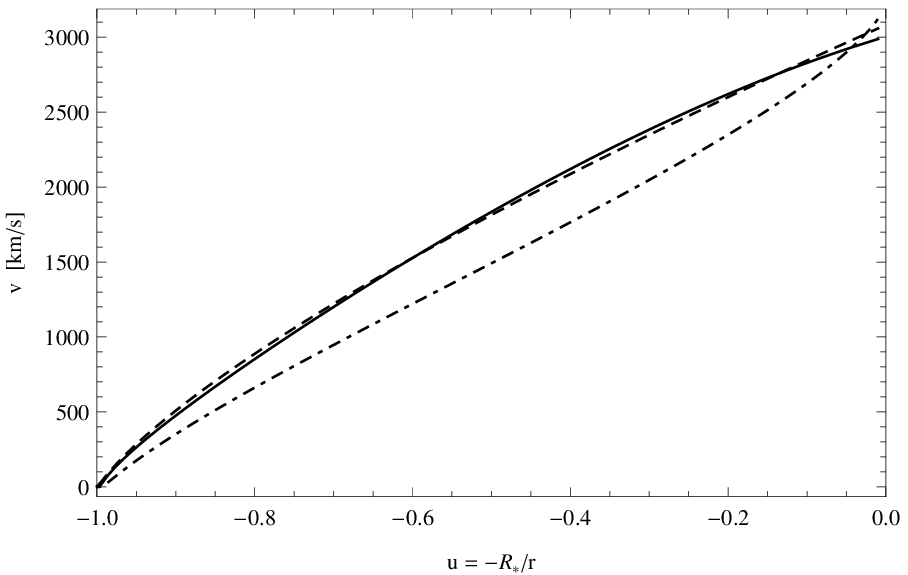}
\caption{Comparison of line accelerations as function of the radial coordinate $\hat{r}=r/R_{*}$ (upper panel) and velocity profile as function of $u\,=\,-R_{*}/r$ (lower panel) with the numerical hydrodynamic solution for an O5\,V star. The line acceleration and the velocity profile calculated using the expressions given by MV08 are shown in dot-dashed line (for the set of parameters: $\hat{g_{0}}=17\,661$, $\delta_{1}=0.6878$, $\hat{r_{0}}=1.0016$ and $\gamma= 0.4758$), those corresponding to \citet{Villata1992} are in dashed--line and the hydrodynamic calculations are in solid line. The analytical expressions derived by MV08 show disagreements in the whole profiles with both \citet{Villata1992} and the numerical calculations, although the agreement in the terminal velocities of the three models can be considered acceptable.
}	
\label{mv08-hydro}
\end{figure}

\begin{figure}[h]
\center
\includegraphics[width=4 in]{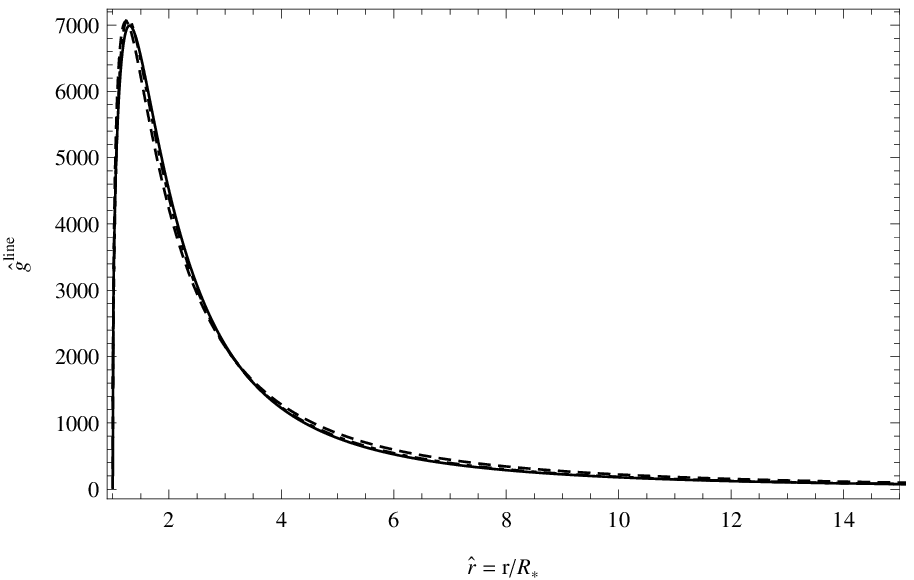}\\
\vspace*{0.4in}
\includegraphics[width=4 in]{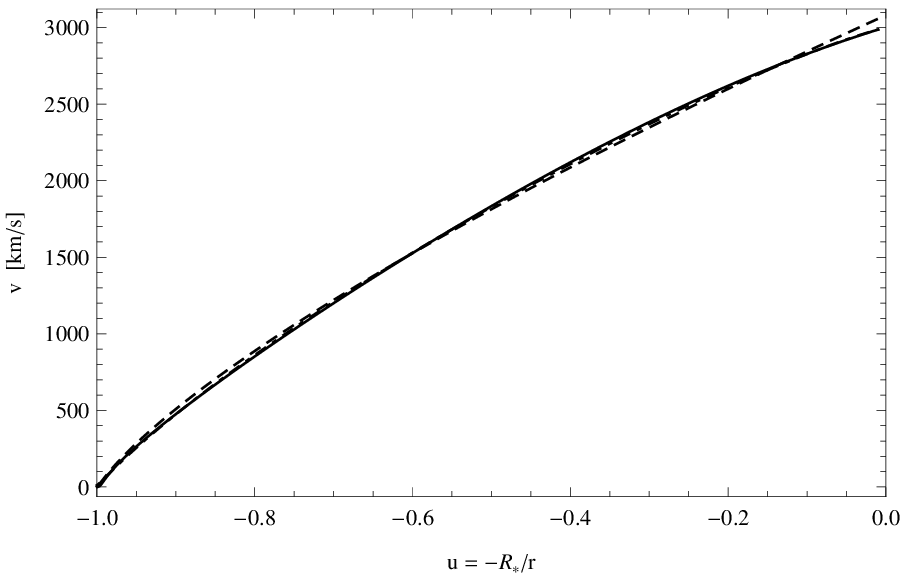}
\caption{Comparison of the analytical line accelerations as function of the radial coordinate $\hat{r}=r/R_{*}$ (upper panel) and the velocity profiles as function of the inverse radial coordinate $u\,=\,-R_{*}/r$ (lower panel) for an O5\,V star, using $\hat{g}^{\rm{line}}_{\mathrm{MV08}}$ (dot-dashed line, with $\hat{g_{0}}=28\,289.8$, $\delta_{1}=1.1585$, $\hat{r_{0}}=0.9926$ and $\gamma= 0.6212$) and $\hat{g}^{\rm{line}}_{\mathrm{V92}}$ (dashed--lines). The agreement among all the results is almost perfect. Solid line corresponds to the numerical hydrodynamic calculations.
}	
\label{mv08-villata-hydro}
\end{figure}

As mentioned previously, the analytical approach proposed by MV08 was calculated \textit{only} for one set of  parameters $\hat{g_{0}}$, $\delta$, $\hat{r_{0}}$ and $\gamma$ that corresponds to an O5 main sequence star with the following stellar parameters: $T_{\rm{eff}}\,=\,40\,000$ K, $R_{*}\,=\,11.757$\,R$_{*}$, $M\,=\,40\,$ M$_{\sun}$ and a solar chemical composition. Therefore, the use of $\hat{g}^{\rm{line}}_{\mathrm{MV08}}$ is very restrictive. Instead, \citet{Villata1992} developed a general expression for the line acceleration, based on the 
standard force multiplier parameters $k$, $\alpha$ and $\delta$, whose values are tabulated in several works \citep[see for example][]{abbott1982,ppk1986,lamers1999}. 

One could wonder if both analytical expressions that describe the behaviour of the radiative acceleration as function of the radius, as well as their corresponding velocity profiles, are similar.    
Fig. \ref{mv08-hydro} displays the comparison of the solutions given by Villata (dashed--line) and  MV08 (dot-dashed line, for the set of parameters given in the iteration A of MV08 work) with the hydrodynamical solution (solid line) calculated for an O5\,V star using the force multipliers parameters listed by \citet[][]{ppk1986}, namely $k=0.124$, $\alpha=0.640$ and $\delta=0.070$.  The results computed with the expressions and parameters given by MV08 present large discrepancies with the numerical hydrodynamical model, both in the behaviour of the line acceleration as function of radius (upper panel) and the resulting velocity profile as function of the inverse coordinate $u=-R_{*}/r=-1/\hat{r}$ (lower panel). Instead, Villata's approximation and the corresponding numerical velocity solution agree very well with the hydrodynamical results, as shown in section \S \ref{compare}.

In principle, we can attribute the discrepancy between Villata and MV08's line--accelerations to the lack of knowledge of a relationship between the force multiplier parameters and the $\hat{g_{0}}$, $\delta_{1}$, $\hat{r_{0}}$ and $\gamma$ parameters. This relationship can be obtained by equalizing $\hat{g}^{\rm{line}}_{\mathrm{MV08}}$ with $\hat{g}^{\rm{line}}_{\mathrm{V92}}$. Then, we obtain,

\begin{equation}
\hat{g}_{0} =\frac{1}{a^{2}}\frac{G\,M\,(1-\Gamma)}{R_{*}}\,A(\alpha, \beta, \delta),
\label{mv-1}
\end{equation}

\begin{equation}
\delta_{1} =1, 
\label{mv-2}
\end{equation}

\begin{equation}
\hat{r}_{0} =1 
\label{mv-3}
\end{equation}

\noindent and

\begin{equation}
\gamma = \gamma_{V} =\alpha \, (2.2 \, \beta -1).
\label{mv-4}
\end{equation}



Based on this,  we find a dependence of $\hat{g_{0}}$, $\delta_{1}$, $\hat{r_{0}}$ and $\gamma$ as a function of force multiplier parameters that agrees very good with Villata's approximation and the numerical hydrodynamical solution. Thus, we show that both expression of the line acceleration are essentially the same approximation and are both based on the m-CAK theory.

Table \ref{tab1} shows the fitted parameters with the values obtained from Eqs. (\ref{mv-1}) to (\ref{mv-4}) and those derived by fitting M\"uller \& Vink's approximation to the $g^{\rm{line}}$ expressed by the m-CAK theory. Using the values given in Table \ref{tab1}, Villata and MV08's line accelerations and velocity profiles are almost identical (see Fig. \ref{mv08-villata-hydro}).

\begin{table}[h]
  \begin{center}
  \tabcolsep 1.2 pt
  \caption{Comparison of $\hat{g_{0}}$, $\delta_{1}$, $\hat{r_{0}}$ and $\gamma$ parameters obtained by fitting a hydrodynamic calculation with those derived using Eqs. (\ref{mv-1}) to (\ref{mv-4}) (Villata's relationship).}
  \label{tab1}
\vskip 0.5cm
{
  \begin{tabular}{l|c|c|c|c}
\hline
\hline
Method  &$\;\hat{g_{0}}\;$ &$\;\delta_{1}\;$ & $\;\hat{r_{0}}\;$ & $\;\gamma\;$\\
\hline
Numerical solution$\;$& $\;28\,289.8\;$  & $\;1.1585\;$& $\;0.9926\;$ & $\;0.6212\;$\\
Villata's relation$\;$& $\;23\,396.6\;$ &  $\;1.0000\;$ & $\;1.0000\;$ & $\;0.4683\;$ \\

\hline
  \end{tabular}
  }
 \end{center}
\end{table}

\subsection{On the new analytical solution}

From the results of Table \ref{sample} we demonstrated our analytical solution based on Villata and MV08 are very good analytical approximations to the radiation driven wind theory. Discrepancies in the terminal velocity and mass loss rate are below $5 \,\%$ and $12\, \%$, respectively, when comparing results from the analytical expression and our numerical hydrodynamical code. In almost all the terminal velocities derived from the analytic solution, we find higher values than those obtained from the numerical hydrodynamics. This could be due to the fact that the velocity rises without limit as the distance increases because of the isothermal temperature assumption. To overcome this problem, MV08 derived an analytical expression for the wind solution in the supersonic approximation by neglecting the corresponding pressure term in the equation of motion, but always in dependence of the $\hat{g_{0}}$, $\delta_{1}$, $\hat{r_{0}}$ and $\gamma$ parameters.

However, as MV08 line force parameters can also be expressed in terms of the force multiplier and stellar parameters (via Eq. 25--28), both formulations will allow us to discuss the scaling laws and test the WML relationship for a large variety of cases. 
In particular, CAK force multiplier parameters required for our analytical solution can be found, e.g.,  in \citet{ppk1986} for stars with temperature over $20\,000$ K. Then, both the velocity profile and the mass--loss rate can easily be obtained without extra computing time. For our case, the mass-loss rate results from the approximation derived by Villata.

At the time \citet{kppa1989} developed their analytic formalism, only one solution of the 
m--CAK equation of motion was known. However, some years later, \citet{cure2004} and \citet{cure2011} reported the existence  of two new hydrodynamical solutions, namely: $\Omega$-slow and $\delta$-slow solutions.  $\Omega$-slow solution corresponds to stars rotating higher than  $\sim$ 75\% of the star critical rotation velocity. On the other hand, it can be shown that the $\delta$-slow solution for early-type stars exists when the $\delta$ line force parameter is greater than 0.25. However, the analytical expressions derived previously (Eq. \ref{villata-gline2}) cannot be applied when the line force parameter is $\delta$ $\gtrsim 0.3$, because of the $\hat{g}^{\rm{line}}_{\mathrm{V92}}$  turns complex.
Therefore, we leave for a forthcoming paper the search for an analytical solution for the equation of motion that can also be used to describe slow radiation-driven winds.

\section{Conclusions}
\label{conclusions}

We obtained an  analytical expression for the velocity profile, solution of the equation of motion for 
radiation--driven winds, in terms of the force multiplier parameters. This analytical expression was obtained by employing the expression of the line acceleration given by \citet{Villata1992} and the methodology proposed by \citet{Muller2008}. We evaluated the accuracy of this new analytical expression for the velocity profile by comparing the mass--loss rate and terminal velocity with numerical 1--D hydrodynamical solutions for radiation-driven winds using the code described in \citet{cure2004}, which we will call hereafter HYDWIND. In all the cases, the analytical expression provided a very good fit with the numerical solution.

In addition, we demonstrated that the terms $\hat{g}^{\rm{line}}_{\mathrm{V92}}$ and $\hat{g}^{\rm{line}}_{\mathrm{MV08}}$ are in essence the same, and both are a very good approximation of the m-CAK theory, and provided useful relationships to determine the parameters for the line acceleration computed by  \citet{Muller2008} in terms of the force multiplier parameters.

The advantages of deriving analytical expressions for the velocity profile is that the resolution of the radiative transfer problem becomes easer. 
Unfortunately, the new analytical solution can only be applied to describe the standard m-CAK theory (a fast wind regime), since for values of $\delta$ \,$\gtrsim 0.3$ the $\hat{g}^{\rm{line}}_{\mathrm{V92}}$ turns complex. A new and more general study for fast rotating stars and winds with ionization gradients will be carried out in a future work.

The existence of various kinds of wind solutions opens the possibility to explore latitudinal--dependent outflowing winds, in which the wind regime might switch from a fast to a slow outflow \citep{cure2005,madura2007}. Therefore, we remark the importance of having analytical expressions to represent the hydrodynamics, mainly if we want to compute the line spectrum in a medium where rotation and the development of inhomogeneities (microclumpings and porosity) also play a fundamental role.

\begin{acknowledgements}

We want to express our acknowledgement to the Referee for his/her comments 
and very helpful suggestions.
This work has been partially supported by FONDECYT project 1130173 and the Centro de Astrof\'isica de Valpara\'iso. IA thanks the support from Gemini-CONICYT 32120033 (Grants for the Astronomy Graduate Program of the Universidad de Valpara\'iso) and Fondo Institucional de Becas FIB-UV. MC also thanks the support from CONICYT, Departamento de Relaciones Internacionales “Programa de Cooperaci\'{o}n Cient\'{i}fica Internacional” CONICYT/MINCYT 2011-656.
LC acknowledges financial support from the Agencia de Promoci\'on Cient\'{\i}fica  y Tecnol\'ogica (Pr\'estamo BID PICT 2011/0885), CONICET (PIP 0300), and the Universidad Nacional de La Plata (Programa de Incentivos G11/109), Argentina. Financial support from the International Cooperation Program MINCYT-CONICYT (project CH/11/03) between Argentina and Chile is also acknowledged.

\end{acknowledgements}

\bibliographystyle{apj}
\bibliography{citas}

\end{document}